\documentstyle[12pt,aaspp4]{article}

\newbox\grsign \setbox\grsign=\hbox{$>$} 
\newdimen\grdimen \grdimen=\ht\grsign
\newbox\simlessbox \newbox\simgreatbox \newbox\simpropbox
\setbox\simgreatbox=\hbox{\raise.5ex\hbox{$>$}\llap
     {\lower.5ex\hbox{$\sim$}}}\ht1=\grdimen\dp1=0pt
\setbox\simlessbox=\hbox{\raise.5ex\hbox{$<$}\llap
     {\lower.5ex\hbox{$\sim$}}}\ht2=\grdimen\dp2=0pt
\setbox\simpropbox=\hbox{\raise.5ex\hbox{$\propto$}\llap
     {\lower.5ex\hbox{$\sim$}}}\ht2=\grdimen\dp2=0pt

\def\simless{\mathrel{\copy\simlessbox}}

\begin{document}

\title{{\it BeppoSAX} Observation of NGC~7582: Constraints on the 
X-ray absorber}

\author {T.J.Turner \altaffilmark{1,2}, G.C. Perola,  \altaffilmark{3}, 
F. Fiore,   \altaffilmark{4,5}, G. Matt,   \altaffilmark{3}, 
I.M. George \altaffilmark{1, 6}, L. Piro,   \altaffilmark{7}, 
L. Bassani \altaffilmark{8}  
}

\altaffiltext{1}{LHEA, Code 660, NASA/Goddard Space Flight Center,
  	Greenbelt, MD 20771}
\altaffiltext{2}{University of Maryland Baltimore County, 1000 Hilltop Circle,
	Baltimore, MD 21250}
\altaffiltext{3}{Dipartimento di Fisica, E. Amaldi,  Universita deglai Studi
      Roma Tre, Via della Vasca Navale 84, I-00146 Roma, Italy} 
\altaffiltext{4}{SAX/SDC Nuova Telespazio, Via Corcolle 19, I-00131 
    Roma, Italy} 
\altaffiltext{5}{Osservatorio Astronomico di Roma, Via dell'Osservatorio, 
   I-00044 Monteporzio-Catone, Italy}
\altaffiltext{6}{Universities Space Research Association}
\altaffiltext{7}{Istituto Tecnologie e Studio Radiazioni Extraterrestri, CNR, Via Gobetti 101, I-40129 Bologna, Italy}
\altaffiltext{8}{Istituto di Astrofisica Spaziale, CNR, Via Fosso del Cavaliere, I-00133 Roma, Italy}


\begin{abstract}

This paper presents a {\it BeppoSAX} observation of NGC~7582 
made during 1998 November and an optical spectrum taken 
in 1998 October. The new X-ray data reveal a previously unknown 
hard X-ray component in NGC~7582, peaking close to 20 keV. 
Rapid variability is observed with correlated 
changes in the 5-10 and 13-60 keV bands  indicating 
that a single continuum component, produced by the active nucleus, 
 provides the dominant flux across both  bands. 
Comparison between  {\it RXTE} and {\it BeppoSAX} data  
reveals   changes in the  
2 -- 10 keV flux on timescales of months. Changes in the nuclear  
X-ray flux  appear unrelated to the gradual decline in optical flux 
noted since the high-state in 1998 July. 
The 0.5 -- 2 keV flux of NGC~7582 is not significantly variable 
within the {\it BeppoSAX} observation, but has brightened by 
a factor of $\sim 2$ since the {\it ASCA} observation of 1994. 
While there is some concern about contamination from 
spatially-unresolved sources, the long-term  variability
in soft X-ray  flux seems most likely  associated with the nucleus 
or an event within the host galaxy of NGC~7582. 

The 2 -- 100 keV spectrum is well fit by a powerlaw of photon 
index $\Gamma=1.95^{+0.09}_{-0.18}$, steeper by 
$\Delta \Gamma \simeq 0.40$  than the 
index during the 1994 {\it ASCA} observation. 
The X-ray continuum is attenuated by a  thick absorber of  
$N_H \sim 1.6 \times 10^{24}{\rm cm^{-2}}$ 
 covering $\sim 60^{+10}_{-14}\%$ of the nucleus plus  a  screen 
with $N_H \sim 1.4 \times  10^{23}{\rm cm^{-2}}$ covering the 
entire nucleus. Comparison of the  
{\it BeppoSAX} and {\it ASCA} spectra shows an increase  
 in the full screen  by 
$\Delta N_H \simeq 7 \times 10^{22} {\rm cm^{-2}}$ since 1994, 
 confirming the absorption variability found by  Xue et al. 
The increase in soft X-ray flux between 1994 and 1998 is consistent   
with the appearance  
of holes in the full screen allowing 
$\simless $ 1\%  of the nuclear flux to escape, and 
producing  some clear lines-of-sight to the broad-line-region. 
The  data are also consistent with the scenario 
suggested by Aretxaga et al, of   
 the radiative onset of a type IIn 
supernova causing the observed optical change in NGC 7582.

\end{abstract} 


\section{Introduction}

NGC~7582 (z=0.0053) 
is known as a Seyfert 2 galaxy and also as a Narrow Emission Line 
Galaxy (NELG). The former classification is made since the 
optical spectrum has historically shown  no broad  
components on the emission lines, nor does spectropolarimetry show any 
broad-line-region visible in scattered light (Heisler, Lumsden 
and Bailey 1997).  The latter classification is based upon 
its brightness in the 2-10 keV band, which led to its detection in early 
hard X-ray surveys (Ward et al. 1978).

An {\it ASCA} observation of NGC~7582 showed a complex spectrum in the 
0.5-10 keV band, composed of a heavily absorbed ($N_H \sim 10^{23} 
{\rm cm^{-2}}$) and flat ($\Gamma \sim 1.4$) continuum, plus 
an unabsorbed and steep component dominating below 2 keV (Turner et al 1997). 
The hard component shows flux variations down to timescales of a few 
thousand seconds (Turner et al. 1997; Schachter et al 1998; Xue et al 1998) 
confirming its association with an active nucleus in this source. 
Xue et al (1998) report a significant change in X-ray spectrum 
between the 1994 and 1996 {\it ASCA} observations of NGC~7582. 
The spectral change was consistent with an increase of the X-ray column by 
$\sim 4 \times 10^{22} {\rm cm^{-2}}$ between these epochs, indicating 
that the X-ray absorber is not a simple uniform screen (Xue et al 1998). 
A  variation in X-ray absorption was previously suggested 
based upon comparison between {\it Einstein}, {\it EXOSAT} and {\it Ginga} 
spectra of NGC~7582 (Warwick et al. 1993). 
The  soft X-ray flux has shown no evidence for 
rapid variability to date. The soft 
flux is most likely composed of contributions from the active nucleus of 
NGC~7582,  from starburst regions or other hot gas in the host galaxy and 
 from  spatially-unresolved nearby  X-ray sources.

Aretxaga et al. (1999; hereafter A99) 
report observations of  
NGC~7582 in a very unusual optical state during July of 1998 when  
the Seyfert showed  
the presence of strong and broad line components blueshifted 
by $\sim 2500\ {\rm km\ s^{-1}}$  relative to the narrow components. 
A99 report the width of H$\alpha$ to be 
FWHM $\sim 12000\ {\rm km\ s^{-1}}$ and the flux state of the source to be 
relatively high at the time of this optical ``event''. 
Following an IAU circular 
announcing an unusual optical state for NGC~7582, we performed 
optical and {\it BeppoSAX} 
and observations in October and November of 1998, respectively.

\section{The {\it BeppoSAX} Observation}

{\it BeppoSAX} (Boella et al. 1997a) 
observed NGC~7582 over the period 1998 November 9 -- 10 
and we report on data from the three co-pointed  narrow-field-instruments:
the Low and Medium Energy Concentrator Spectrometers (LECS and MECS, 
respectively), covering 
$\sim 0.1$ -- 10 keV (Parmar et al. 1997) and 1.3 -- 10 keV 
(Boella et al. 1997b) and the Phoswich Detector System 
(PDS; Frontera et al. 1997) providing data over $\sim 13$ -- 200 keV. 
The full-band of each instrument can be used for construction of 
light curves and images, but in the spectral analysis (\S2.3) only 
a subset of data are used; 0.1 -- 4 keV (LECS), 1.5 -- 10 keV (MECS) 
and 15 -- 100 keV (PDS). 
This observation was part of a large spectral survey program, but 
was brought forward to occur as close as possible to the unusual optical 
event. 

Standard data reduction was performed using the SAXDAS software 
package version 2.0 following Fiore, Guainazzi \& Grandi (1999). 
Data from MECS2 and MECS3 were combined, as were data from the four PDS 
phoswich units,  after gain equalization and linearization. 
The instruments are switched off during SAA passages and 
PDS data acquired during the first 5 minutes after each SAA 
passage have been rejected to exclude periods of highly 
variable risetime threshold (see Fiore, Guainazzi \& Grandi 1999). 
This exclusion reduces the total 13-200 keV 
background from 30 to about 20 ct s$^{-1}$ and the 13-80 keV 
background from 20 to about 12 ct s$^{-1}$. 
The PDS rocking mode 
provides a very reliable background subtraction. This has been 
checked using the spectrum between 220 and 300 keV, 
where the effective area of the 
PDS to X-ray photons is negligible and therefore the source 
contribution is negligible. 
We obtain, after background subtraction,  
0.014$\pm0.012$ counts s$^{-1}$ consistent with the expected value of 0.

Standard screening of the {\it BeppoSAX} data  yielded 
events files  with effective exposure times of 
27.3 ks (LECS), 56.4 ks for the combined MECS2/3 instruments and 
52.2  ks for the PDS. The LECS exposure is lower than that in the 
MECS because the former is switched off during periods when the Earth's 
illuminated limb is close to the line-of-sight.

\subsection{Contamination Issues}

The LECS and MECS instruments have fields-of-view ($fov$) with radii  
$22'$ and $28'$ respectively, while the PDS has a collimated 
$fov$ with FWHM = 1.4 degrees. The region around NGC~7582 
contains numerous  X-ray sources evident in the {\it BeppoSAX } data 
(Fig.~1). 
Catalogues available online (via the HEASARC) show
17 AGN exist within a  1.5 degree radius of NGC~7582, in addition to the  
cluster Sersic159-03. Of the AGN, the  BL Lac PKS 2316-423 and 
the Seyfert NGC 7590 are brightest in the soft X-ray band. 
However, most of the sources in this field have a soft 
X-ray spectrum, as is evident by comparison of the left and right panels 
of Fig.~1. In fact, 
 construction of a MECS  image in the 
5-10 keV range allows  confirmation 
 that none of the known AGN within the 
 $fov$ are detected above 5 keV, nor is there any other hard X-ray 
source of comparable flux to NGC~7582 within the MECS $fov$.   There
is a chance that a significant hard source may exist outside of the MECS 
$fov$ but within the PDS collimator; however, there is no such known source. 

As reported by Schachter et al (1998), there is a significant  
unidentified soft X-ray source detected 
at J2000 $23^h 18^m 29.9^s$, $-42^{\circ} 20' 41.4''$ in 
 {\it ROSAT}  HRI and PSPC images,  
i.e. about $2'$ from NGC~7582 and which we henceforth refer to as 
RX J231829.9-422041. During the 17.2 ks HRI observation 
Schachter et al (1998) found 
$225^+_-32$ counts from NGC~7582 and $90^+_-26$ counts from 
RX J231829.9-422041.

The proximity of RX J231829.9-422041 would be irrelevant if we could 
spatially resolve this source from NGC~7582 in the LECS and MECS data. 
The MECS have a  spatial 
resolution of $1.2'$ (FWHM) at 6 keV, allowing the potential separation 
of any hard X-rays from the serendipitous source and NGC~7582. However 
 the spatial resolution of the  
LECS is $3.5'$ at 0.25 keV, making separation difficult at the softest 
energies. 
Examination the LECS and MECS images (Fig.~1) shows evidence for 
a significant region of unresolved soft X-ray emission  on 
the north-east side of NGC~7582, consistent with the position 
of  RX J231829.9-422041. Unfortunately the 
 LECS emission from RX J231829.9-422041 is not separable from the 
tail of the point-spread-function  of LECS counts from  NGC~7582. 
Thus we investigate the nature of 
RX J231829.9-422041, to determine whether we can quantify the 
contamination of the NGC~7582 data based on the {\it ROSAT} flux and 
spectrum. 

Overlaying optical and X-ray images 
shows that  RX J231829.9-422041 appears to lie outside of the host galaxy 
(which has a $\sim 5' \times 2.5'$ diameter on the sky, from 
the {\sc skyview} digitized Southern Sky Survey plus Palomar 
Sky Survey E plates) 
 and hence may be a foreground source or background quasar. 
We obtained a spectrum of an optical candidate at 
J2000 $23^h18^m29.94^s$, $-42^{\circ} 20' 40.7''$ 
 using the 4m telescope 
at CTIO with RC spectrograph and Loral 3K CCD. 
The data were accumulated with a spectral resolution of 6.3 \AA\ over
$\sim 4000-9000$ \AA\ and a WG360 blocking filter was used. 
Wavelength calibration utilized a helium/argon 
spectrum yielding an accuracy of 3 \AA. 
The data were extracted using standard methods and the 
IRAF software. The optical spectrum of 
RX J231829.9-422041 was flux-calibrated using standard star LTT9491  
(Stone \& Baldwin 1983; Baldwin \& Stone 1984); target and 
standard star were observed close to the zenith using a $1.5''$ 
slit and under photometric conditions. 

The spectrum was 
accumulated UT 1998 October 17 and we show the data binned to 
a resolution of 8  \AA\ 
over the $\sim 4500$ -- 9000 \AA\ bandpass (Fig.~2). 
An unresolved line is evident at 6593 \AA\ and 
a broad line at 8638  \AA\ . The presence of 
a strong broad line (FWHM width $\sim 3000 {\rm km\ s^{-1}}$) 
indicates the source is  an 
 AGN. Assuming the broad line is 
H$\alpha$ then the redshift is z=0.316 and the 
 6593 \AA\ line is consistent with 
 \verb+[+ O{\sc iii} \verb+]+ $\lambda 5007$.   
H$\beta$ is expected at 6397 \AA\ but not observed, this could be relatively 
weak and hidden in the noise. 
However, Schachter et al (1998) note another, fainter, 
 optical candidate within the 
HRI error box, for which we have no optical spectrum, 
thus the identification of the 
X-ray source remains inconclusive. 

 We compared the X-ray flux of RX J231829.9-422041  
 during the PSPC and HRI observations and found 
$F_{0.1-2}=8.7^+_-1.0 \times 10^{-14}$ and 
$1.4^+_-0.40 \times 10^{-13} {\rm erg\ cm^{-2}s^{-1}}$ 
respectively, assuming the model noted below.
A 58\% increase is seen between the two observations, which are 
separated by $\sim 2 $ years, 
consistent with the tentative AGN identification. If the optical 
source is the correct counterpart of RX J231829.9-422041, then for a 
redshift z=0.316, the luminosity is 
$L_{0.1-2}=3.6 \times 10^{43} {\rm erg\ s^{-1}}$.

Thus we conclude RX J231829.9-422041 is likely to be an AGN, and 
the source variability makes contamination in the LECS data a 
problem. To minimize contamination we choose a special extraction cell
for the LECS data from NGC~7582. 
The standard LECS extraction cells of $8'$ or $4'$ 
would include some contaminating emission from these sources, as is evident 
from Fig.~1. To minimize contamination of the soft spectrum we 
choose a LECS extraction cell using  sky coordinates 
which allows selection of spectral 
 counts only from the south-west semi-circle of a 
circular region centered on the nucleus of NGC~7582. The position angle for 
the dividing line was 128 degrees, measured anti-clockwise from North. 
This was  properly scaled 
to reflect its contribution of only half the normal LECS flux, 
and  provides the least contaminated LECS spectrum possible.

\subsection{Timing Analysis}

Fig.~3 shows the light curves of NGC~7582 sampled with 5000 s bins, 
in different bandpasses. The LECS on-source 
light curve was extracted from the semi-circular region described above,
and hence represents only half of the  LECS count rate of the source 
in a standard 8' circle in  
the 0.1--1 keV bandpass. The LECS background light curve is also shown.
No significant variability is detected in the 0.1--1 keV band 
($\chi^2=17.9/17\ dof$). 
The  MECS light curve shows short-term variability in the 
5-10 keV flux, with factor of two variations on timescales of 
$\sim 15,000$s and lower amplitude variability occurring on timescales 
of a few thousand seconds. The MECS background level is negligible 
and not shown.  The PDS light curve is
shown in the 13-60 keV band, and appears consistent with the MECS 
light curve (although we are not sensitive to small lags). 
This indicates that the PDS data for NGC~7582 are not significantly 
contaminated by any other source. Furthermore, 
the similarity in 5-10 and 13-60 keV light curves  indicates that 
the emission in those bands originates in the same physical component, or in very 
closely linked regions (see next section). 

NGC~7582 yielded 0.013$^+_-0.008$ source ct/s in the LECS (0.1 -- 4 keV band; 
this is the rate in the half-cell so 
should be doubled for comparison 
with other sources), 
0.166$^+_-0.002$ ct/s in the MECS (1.5 -- 10 keV) and 
0.82$^+_-0.04$ ct/s in the PDS (15 -- 100 keV) corresponding to 
fluxes $F_{0.5-2}=5.66 \times 10^{-13} {\rm erg\ cm^{-2}s^{-1}}$, 
  $F_{2-10}=1.97 \times 10^{-11} {\rm erg\ cm^{-2}s^{-1}}$, 
 $F_{10-100}=1.16 \times 10^{-10} {\rm erg\ cm^{-2}s^{-1}}$ for
the best-fitting model (see \S 2.3). Uncertainties in absolute flux 
calibration are $\sim 15\%$.

\subsection{Spectral Analysis}

Spectral data were extracted from a circular region of radius 
 $4'$ for the MECS, chosen to encircle 
$\sim 90\%$ of the source counts. The LECS data were taken from a co-centered 
semi-circle of radius $8'$, as described in detail in \S2.1.  
LECS and MECS background spectra were taken from blank fields, using the same 
region of the detector in each case. It has been determined that
this method of extracting background spectra yields a superior 
result to that obtained using an offset region from the source events file  
because the LECS and MECS background strongly varies with position in 
the detector.

In this section the {\it BeppoSAX} data were fit with the normalization 
of the MECS dataset fixed at 1; the PDS data were 
allowed a normalization in the range 0.75-0.85 
(0.8 is the expected value found from fitting calibration 
sources). The LECS normalization was allowed to vary freely and 
was usually close to 0.45 (usually 0.9, but we have only half the source flux 
as noted above). 
Corrections for the source flux falling outside of the 
extraction cell due to the instrument point-spread-function 
 were incorporated into the {\sc arf} files 
 used in the spectral fitting. 

\subsubsection{The 3-100 keV  Continuum}

First, the {\it BeppoSAX} data were fit in the limited range 
of 3-5 keV plus 7-100 keV, using a single absorbed powerlaw 
model. This was to ascertain whether the continuum 
could be modeled in a simple way, when the complex 
soft emission and the iron K$\alpha$ line band were excluded. 
This model proved inadequate yielding $\chi^2=101/60\ dof$ and 
$\Gamma \sim 1.46$. 
The ratio of data/model (Fig.~4) shows data  in the PDS band 
do not lie on the extrapolation of the continuum fitted to 
data from below 10 keV,  suggesting 
the presence of an additional spectral  component above 10 keV.

There are several models which can be tried.  
First, a complex absorber was fit to the data.
The  absorber component of the model 
was changed  to a ``dual-absorber'', whereby lines-of-sight to the continuum 
pass through one or both absorbers. 
This model yielded a good fit 
with $\chi^2=59/58\ dof$. The data are described by a 
continuum of photon index $\Gamma=1.95 $ of which 60\% is covered 
with a column $N_H=1.60^{+0.94}_{-0.46} \times 10^{24} {\rm cm^{-2}}$ 
and 100\% is covered by 1.44$^{+0.09}_{-0.10} \times 10^{23}{\rm cm^{-2}}$;
hereafter these absorbing components are referred to as the ``thick absorber''
and the ``full screen'', respectively. 

Some alternative parameterizations were tried for the hard spectral component.
A reflection component was added to the model in place 
of the heavily absorbed fraction of powerlaw.
The  {\sc xspec pexrav} 
model was used, representing reflection from neutral material (here 
assumed to be viewed at an angle 
$cos \theta = 0.45$  to the line-of-sight) of Magdziarz \& Zdziarski 1995. 
First the reflection component was assumed to be attenuated 
by the same column affecting the continuum, yielding 
$\chi^2=52/59\ dof$ and a  solid angle of 
$\Omega/2 \pi=6.0$ steradians. If the PDS data were dominated by 
reprocessed emission from a geometrically large reflector then 
the PDS light curve would be expected 
to be smeared and lagged relative to the primary light curve.
However, such large solid angles 
may have no geometrical meaning but may indicate 
anisotropy of the nuclear radiation. A test of the latter model 
is the consistency between the strengths of the iron line and 
reflection hump. In fact the reflection component is 
inconsistent (at $> 99\%$ confidence) 
with the strength of the iron K$\alpha$ line 
(George \& Fabian, 1991). We reach the same conclusion if the 
 reflection component suffers no absorption (except Galactic, 
$N_H=1.47 \times 10^{20} {\rm cm^{-2}}$, Stark et al. 1992). 
We conclude the hard component in NGC~7582 is not dominated by 
Compton reflection of the primary continuum. 

We also  tried a model where the hard X-ray spectrum suffers an  
exponential cutoff, to explain the spectral shape above 10 keV. 
When left unconstrained  the cutoff energy reaches a solution 
beyond the upper limit of the PDS bandpass,  rejecting the 
cutoff as a significant model parameter. When the 
cutoff energy was constrained to match the apparent turnover 
evident in the data (around a few tens 
of keV)  then the continuum powerlaw was found to be very  flat 
$\Gamma \sim 1$ and the fit was  inferior 
($\chi^2=101/59\ dof$) to that found for the complex absorber model  
described above. Thus the dual-absorber model is favored over 
an exponentially cutoff powerlaw.

\subsubsection{The soft X-ray continuum}

In the previous section we found the 
 best description of the 3-100 keV continuum to be  
a $\Gamma=1.95$ powerlaw, 60\% of which is absorbed by 
a column $N_H=1.60 \times 10^{24}{\rm cm^{-2}}$ (the thick absorber) and 
100\% by $1.44 \times 10^{23} {\rm cm^{-2}}$ (the full screen).  
Extrapolation of this model below 3 keV reveals the soft X-ray data 
to be in excess of the model, and underlines the requirement 
for an additional component to parameterize the soft emission. 
To this end, a  {\sc mekal} component 
(assuming cosmic abundances) was added to the dual-absorber model
detailed above, absorbed by a column fixed at the 
Galactic line-of-sight value 
$N_H=1.42 \times 10^{20} \ {\rm cm^{-2}}$. 
{\sc mekal} is a model for the  emission spectrum from 
collisionally ionized gas (Kaastra, 1992 and references 
within). At this stage, the iron line 
band of 5.0-7.0 keV remained excluded. 

This model did not adequately describe all of the soft emission, 
yielding $\chi^2=153/134\ dof$, and so 
an additional component was added, representing an unabsorbed fraction of 
the hard X-ray continuum powerlaw. 
The  soft flux can be attributed to a 
 {\sc mekal} component of temperature $kT=0.72^{+0.29}_{-0.15}$keV.  
The unabsorbed powerlaw comprises $\sim$ 0.4\% of the continuum 
(the so-called full screen may cover only 99.6\% of the nucleus,
 or this flux may arise outside of the full screen). 
In this fit $\chi^2=139/132\ dof$ 
and  the {\sc mekal} component accounts for $41\%$ of 
the total 0.1 -- 2  keV flux (Fig.~5). 
 The total luminosity in the 
0.1 -- 2 keV band is $L_{0.1-2} \sim 4 \times 10^{40} {\rm erg\ s^{-1}}$ 
 at the redshift of NGC~7582. 
 A second {\sc mekal} component does not provide an adequate alternative  
to the  ``leaking'' unabsorbed powerlaw component.

\begin{deluxetable}{lcccccc}

\tablecaption{ {\it BeppoSAX} and {\it ASCA}  Parameters}

\tablehead{
\colhead{} & 
\colhead{ Flux$_{0.5-2}$ } &
\colhead{ Flux$_{2-10}$ } & 
\colhead{ $\Gamma$ } &
 \colhead{ $N_{H}^{FS\ a} $ } &
\colhead{ $N_H^{TA\ b}$ } & 
 \colhead{ $\chi^2 /dof$ } \\
\colhead{} & 
\colhead{ $10^{-13} $ } & 
\colhead{ $10^{-11} $ } & 
\colhead{} & 
\colhead{ $10^{23}{\rm cm^{-2}}$ } & 
\colhead{ $10^{23}{\rm cm^{-2}}$ }  
&  \colhead{} \\
 \colhead{} & 
\colhead{ ${\rm erg\ cm^{-2}s^{-1}}$ } & 
\colhead{ $ {\rm erg\ cm^{-2}s^{-1}}$ } & 
\colhead{} & 
\colhead{  } & 
\colhead{ }  
&  \colhead{}
}
\startdata

\hline
SAX & $5.66^+_-0.85$ & $1.97^+_-0.3$ & 1.95$^{+0.09}_{-0.18}$ & 
	$1.44^{+0.09}_{-0.10}$ & 
	16.00$^{+9.40}_{-4.60}$  & 59/58 \nl
\hline

A94$^c$ & 3.63$^{+0.49}_{-0.67}$ & 1.51$^{+0.36}_{-0.42}$ &   1.33$^+_-0.16$ &
	 0.75$^{+0.05}_{-0.04}$ & \nodata &  300/288 \nl 

\hline

A96$^c$  & 4.18$^{+0.34}_{-0.40}$ & 1.55$^{+0.24}_{-0.22}$ & 
	$1.52^+_-0.13$ & 1.20$^{+0.11}_{-0.08}$ 
	& \nodata &  601/578 \nl  

\hline
\tablenotetext{a} {FS is the full screen absorber completely  covering  the 
nucleus} 
\tablenotetext{b} {TA is the thick absorber covering $60^{+10}_{-14}\%$ 
of the nucleus}
\tablenotetext{c} {{\it ASCA} results from 1994 and 1996 taken from 
Xue et al. 1998, from their Table 1, fit 2} 

\enddata
\end{deluxetable}

\subsubsection{The iron K$\alpha$ line}

With the  acceptable dual-absorber model, we returned the 5.0-7.0 keV 
data to the spectral fit and turned to an examination 
of the  iron K$\alpha$ line. If the model allowed 
an unconstrained line width then the line went to the maximum 
allowed width of 1 keV. 
The line can be difficult to parameterize if the continuum 
is complex, and we considered the possibility that we still have not 
achieved a correct absorption model for this source,  in which case 
the source might show some unmodeled absorption from the K edge of iron. 
It is important to establish whether we have adequately 
parameterized the absorber because otherwise confusion arises 
between which are line and continuum photons. 
However, addition of an edge to the model 
did not significantly improve the fit. 

Given that the line is quite faint, we tried 
fits fixing the width at values of $\sigma=0.41$ keV 
 (to compare with the {\it ASCA} fit), 0.0 keV (to examine the 
parameterization as a narrow line) and 
the intermediate value of 0.2 keV 
(to expand our examination of parameter-space).
 With $\sigma=0.41$ the fit yields a line 
at rest-energy $6.15^{+0.50}_{-0.65}$ keV and  $\chi^2=149/150 \ dof$. 
The data and residuals for the dual-absorber fit including 
this iron K$\alpha$ line  are shown in Fig.~5, along with the model.  
If we assume the 
line and continuum are both absorbed then the line equivalent width is  
 EW$=186^{+114}_{-79}$eV (absorption-corrected line against 
absorption-corrected continuum).  With $\sigma=0.2$ we obtain a 
rest-energy $6.24^{+0.29}_{-0.33}$ keV, EW$=125^+_-77$ eV
and $\chi^2=153/150\ dof$. 
If the line is assumed to be 
narrow ($\sigma=0$) then we obtain EW$=90^{+57}_{-40}$eV
and $\chi^2=157/150\ dof$.
 Alternatively, if we assume the iron line is not absorbed, then
the narrow Gaussian fit produces EW$=570^{+570}_{-120}$eV against 
the absorbed continuum level. In realistic physical models the 
line is produced within the absorbing clouds, not in front of or behind them 
as in our  parameterizations. 
Despite this shortcoming we can make a crude comparison with  
the  line EW  predicted by  transmission of X-rays through a  
 sphere of the thick absorber, which  would be several keV 
(Leahy and Creighton, 1993). The full screen of gas should produce 
a narrow line of $\sim 100-150$ eV. 
However, the preference for a broad line in these data suggests
 relativistic effects may be important, inferring an origin for the line 
close to the black hole.

\subsection{Comparison with the ASCA data}

The {\it BeppoSAX} data were compared  with the 
best-fitting double-powerlaw solution found for the 
{\it ASCA} 1994 data (with hard photon index 
$\Gamma=1.38$, absorbed by column $N_H =7.4 \times 10^{22} {\rm cm^{-2}}$, 
plus soft unabsorbed component parameterized by $\Gamma= 2.68$;  
Turner et al. 1997). First 
the {\it ASCA} model was compared to the data 
with no refitting or re-normalization, and the data/model ratio 
 is shown in Fig.~6a.
Clearly the AGN has brightened in the hard and soft bands since 1994 
with evidence for a larger absorption 
at the {\it BeppoSAX} epoch. 
Our choice of extraction region for the LECS should reduce contamination 
of NGC~7582 by the north-east sources and so this change in soft flux 
is most likely closely associated with the nucleus. The 
 {\it ASCA} data were probably contaminated by  
significant flux from RX J231829.9-422041. However, the fact the 
uncontaminated soft flux measurement of NGC~7582 by {\it BeppoSAX} 
is {\it higher} than the {\it ASCA} measurement 
means a significant soft flux increase must have occurred in NGC~7582. 

There is a dip at 6 keV but this  does not represent 
a  significant difference in the flux of the iron K$\alpha$ line 
because the {\it ASCA} 1994 measurement of this line (Turner et al 1997) 
 had an uncertainty of 
a factor $\sim 2$. Fig~6b compares the {\it BeppoSAX} spectra with
the second {\it ASCA} observation, from 1996, finding NGC~7582 brighter 
during the 1998 {\it BeppoSAX} observation. 
Fig.~6c shows a comparison between the two 
{\it ASCA} observations.  
Significant 
changes in absorbing column   are 
evident between the two {\it ASCA} observations, as reported in detail by
Xue et al. (1998) and previously noted in this source 
(Warwick et al. 1993), similarly, column changes are 
evident between the {\it ASCA} and {\it BeppoSAX} epochs. 
Of course, the effect of a column 
as large as $\sim 10^{24}{\rm cm^{-2}}$ is 
not detectable below 10 keV and the  column variations 
evident in the {\it ASCA-ASCA} and {\it ASCA-BeppoSAX} 
comparisons occur in the full screen. 
  
Flux variability is also observed on long and short timescales 
in the hard band, and 
on long timescales in the soft  X-ray band. A limitation of plots such as 
Fig.~6 is the absence of an indication of the uncertainty in 
the {\it ASCA} model, to which we compare the {\it BeppoSAX} data. Thus we 
provide the ratio of the {\it ASCA} 1994 data to the best-fitting 
 model to those 1994 data (Fig.~6d) 
to give an idea of the (small) residual uncertainty in the {\it ASCA} fit. 
To conclude, {\it BeppoSAX} data provide evidence for 
a more complex spectrum than previously known for this source; furthermore,  
significant flux, column (full screen) and index variability are evident on 
timescales of years. 

Next we tried to parameterize the long-term spectral
changes by application of the {\it BeppoSAX} dual-absorber model 
to the  1994 November {\it ASCA} data. 
Data from 1994 appear most dramatically different to the spectrum at the 
{\it BeppoSAX} epoch and, as the differences between the two 
{\it ASCA} observations are discussed by Xue et al. (1998), we 
concentrate on this 1998 to 1994 comparison.
 
\subsubsection{Comparison of X-ray Spectra from 1994 and 1998}

Fitting the {\it ASCA} data from the 1994 observation with 
the BeppoSAX model we found that 
 the spectral variability between 1998 and 1994 
 could not be explained by a simple change of one 
spectral parameter. Further analysis showed  it could not 
be understood by allowing the column densities  
of the thick absorber and/or full screen to vary even if 
both covering fractions were also free.  

As the spectral variability 
appears complex, the {\it ASCA} data were refit with the BeppoSAX 
model, this time allowing all spectral parameters to vary.  
This fit yielded 
a spectral  index of $\Gamma=1.55^{+0.19}_{-0.15}$, column of 
$N_H=7.64^{+0.86}_{-0.68} \times 10^{22} {\rm cm^{-2}}$. 
These values agree with similar  fits to  
the {\it ASCA} data obtained by Turner et al. (1997).  
Thus the photon index appears to have steepened by 
$\Delta \Gamma = 0.40$ between 1994 and 1998, while the 
full screen has increased by 
$\Delta N_H \sim 7 \times 10^{22} {\rm cm^{-2}}$. 
The fit gave $\chi^2=509/392\  dof$, with no systematic 
residual. 

The change in soft flux can be modeled as an increase in normalization 
of the {\sc mekal} component by a factor 2.2$^{+1.2}_{-0.6}$. 
The unabsorbed fraction of powerlaw was consistent with 
that found at the {\it BeppoSAX} epoch, as was the {\sc mekal} temperature,
and all parameters of the iron K$\alpha$ line (albeit poorly 
constrained). 

We considered whether the 
full screen might be partially-ionized, 
however such models found no better fit 
than those presented here, and given the difficulty in separating nuclear and 
starburst components below 2 keV, we do not pursue this further.
It is evident that significant flux and column changes occur 
closely associated with the nucleus of NGC~7582, but the 
spectral changes are complex.

\section{The Optical Spectrum}

An optical spectrum of NGC~7582 was obtained on UT 1998 October 17 
(Fig.~7) using
the CTIO 4m telescope with the setup as described in \S2.1. 
 NGC~7582 was flux-calibrated using 
standard star LTT9491 (Stone \& Baldwin 1983; Baldwin \& Stone, 1984) both 
of which were 
observed close to the zenith using a 1.5$''$ slit, under 
photometric conditions. NGC~7582 was observed for a total 
exposure time 1200 s. 
The 1998 Oct 17 spectrum shows a broad component on H$\alpha$, 
with width FWHM=6400 ${\rm km\ s^{-1}}$, and a peak indicating a redshift of 
$\sim 590\ {\rm km\ s^{-1}}$ 
relative to the narrow lines. We find our line widths in good 
agreement with those found by A99 for 1998 Oct 21, and 
concur that at this epoch, the broad line showed a slight redshift 
relative to the narrow lines.

\section{Discussion}

Interest in the {\it BeppoSAX} X-ray spectrum is twofold; providing 
spectral constraints on NGC~7582 up to 100 keV, and providing an 
 X-ray spectrum which offers clues to the nature of a 
significant and unusual optical event associated with the nuclear 
regions. 

\subsection{The X-ray Results}

First we summarize the new  X-ray results. 
{\it BeppoSAX} reveals a hard X-ray component, peaking at $\sim 20$ keV, 
previously unknown in this source. 
By now, {\it BeppoSAX} has discovered numerous examples of continuum components
visible only above 10 keV, 
for example NGC~1068 (Matt et al. 1997); 
NGC~3393 and NGC~4941 (Salvati et al. 1997); Mrk~3 (Cappi et al 1999); 
NGC~2110 (Malaguti et al 1999) and Circinus (Matt et al. 1999). 
In some cases the hard component has been found to be consistent with 
Compton Reflection of the continuum while
in others a heavily absorbed 
($N_H >  10^{24}{\rm cm^{-2}}$) 
fraction of the primary continuum is 
suggested. For NGC~7582 we prefer the latter model. 
We find variability  in the flux above 10 keV 
 on timescales down to a few thousand seconds, 
correlated  with that seen below 10  keV.  Combined 
with the spectral result this suggests that  
a powerlaw continuum of photon index $\Gamma=1.95$ 
dominates from 2 -- 100 keV. This continuum is 
transmitted through a ``thick absorber'' with  
a column $N_H=1.60^{+0.94}_{-0.46} \times 10^{24}{\rm cm^{-2}}$ 
covering 60\% of the nucleus plus a ``full screen'' covering 
100\% of the source with  
$N_H=1.44^{+0.09}_{-0.10} \times 10^{23}{\rm cm^{-2}}$. 
The {\it BeppoSAX} model for NGC 7582 can be extrapolated to 150 keV 
and compared with the OSSE measurement of flux 
(taken from data summed from observations spanning 1991-1994). We find 
for {\it BeppoSAX} 
$F_{50-150\ keV}=6.88 \times 10^{-11} {\rm erg\ cm^{-2}s^{-1}}$ 
and for OSSE 
$F_{50-150\ keV}=3.84^{+1.06}_{-0.56} \times 10^{-11} 
{\rm erg\ cm^{-2}s^{-1}}$ (Johnson et al 1994) 
thus BeppoSAX finds NGC~7582 significantly brighter 
($\sim 80\%$) than the average flux-state sampled by  OSSE data 
summed from observations in 1991, 1992 and 1994.  Alternatively 
the spectrum may cut-off sharply above 50 keV, leading to an overestimate 
of flux when extrapolating the BeppoSAX spectrum up to 150 keV.

In NGC~7582 {\it BeppoSAX} data provide some insight into the 
 geometry of the X-ray absorber. There are several models which would 
allow observation of a nuclear source through differing column densities. 
However, a  single absorber which is thick at the center and tapered 
towards the edges would 
produce a smooth range of columns from center to edge. This would look like a 
flat spectrum, rather than show the relatively sharp turn-over seen in 
the 5 -- 10 keV band (Fig.~5).  
 A similar  spectrum is expected from a nuclear 
source viewed through a distribution of many clouds of varying 
column densities. Thus the data provide strong evidence against those 
geometries. The measured columns and fractions 
 are consistent with a thick absorber composed of a single cloud, or 
several clouds with  
$N_H \sim 1.6 \times 10^{24}{\rm cm^{-2}}$ 
 covering $\sim 60\%$ of the nuclear source. The screen 
 with $N_H \sim 1.4 \times  10^{23}{\rm cm^{-2}}$ 
fully covers the source. The location of the full screen relative 
to the thick cloud(s) is unknown. 
 The thick absorber will provide an 
optical depth of $\sim 1$ to electron scattering. Any quasi-spherical 
 distribution of such 
clouds will result in the introduction  of smearing and lags 
of the intrinsic  variations in the nuclear light 
curve. However, the absence of very sharp flux variations, and the 
uncertainty as to the intrinsic light curve do not allow us 
to place any constraints on the basis of the flux variability
observed. 

The absorber in NGC~7582 can be compared to the complex 
absorbers found in some other Seyfert galaxies. 
NGC~2110 (Malaguti et al. 1999) and 
IRAS 04575-7537 (Vignali et al. 1998) show 30\% covering by 
a column $\sim few\  \times 10^{23} {\rm cm^{-2}}$ plus full-covering   
by   $\sim few\   \times 10^{22} {\rm cm^{-2}}$ and so 
in comparison NGC~7582 has an order of magnitude higher column 
in each component. Seyfert~1 galaxies may have a similarly complex 
geometry of circumnuclear absorption, but with  gas of much higher ionization
state. Several Seyfert~1 galaxies have shown evidence for two zones of 
absorbing gas with differing ionization-state (e.g. Kriss et al. 1996;
George et al. 1998).

Comparison with previous {\it ASCA} 
observations allows us to look for changes in the full screen. 
Application of the {\it BeppoSAX} model to  the 
{\it ASCA} data  suggest a steepening of the photon index 
by $\Delta \Gamma=0.40$ and an increase of column 
by $\Delta N_H \sim 7 \times 10^{22} {\rm cm^{-2}}$ between 1994 and 1998. 
If changes in column and index are 
connected, then changes in the  circumnuclear absorption 
 could be linked to  changes in the region which  Compton 
upscatters nuclear photons.  
As discussed by Xue et al. (1998), assuming column changes 
are caused by transverse passage of imhomogeneities in the absorber 
crossing the line-of-sight, then  the minimum timescale for
column variations (i.e. that found between the two {\it ASCA} 
observations) can be used to obtain a limit to the radial distance  
from the nucleus  and clump
sizes within the absorber. Xue et al. (1998) conclude the full screen 
to be  consistent with the putative torus. 

It is unclear whether any small fraction of the  nuclear continuum 
escapes without attenuation,  the unabsorbed soft X-rays that we observe 
could arise from starburst regions outside of the absorber. The X-ray emission
from starburst and AGN cannot be separated spatially or spectroscopically 
using these X-ray data.

\subsection{The relation between Optical and X-ray Properties}

Now we examine the {\it BeppoSAX} results in light of the recent 
 optical brightening and 
the development of broad components to some permitted lines sometime 
between  1998 June 20 
(Halpern, Kay \& Leighly 1998) and 1998 July 11 (A99). 
The constant optical centroid  
of NGC~7582 (within an error radius of 12 pc) indicates that this
``event'' was closely associated with the nucleus  (A99). 
  Our optical spectrum is in agreement with October 1998 
spectra shown by A99, confirming their result that   
NGC~7582 had returned to close to its typical ``type 2'' characteristics 
by 1998 October,  
although with some  obvious broad component remaining on H$\alpha$. 

To review the recent X-ray flux history of NGC~7582: 
{\it ASCA} observations showed $F_{2-10} = 1.51 \times 10^{-11}$ 
and $1.82 \times 10^{-11} {\rm erg\ cm^{-2}s^{-1}}$ during 
1994 November and 1996 November, respectively. A {\it RXTE} observation 
during 1998 October revealed $F_{2-10}=9.5 \times 10^{-12} 
{\rm erg\ cm^{-2}s^{-1}}$ (Takeshima, p.comm). This {\it BeppoSAX} observation 
shows $F_{2-10}=1.97 \times 10^{-11} {\rm erg\ cm^{-2}s^{-1}}$. 
We conclude that NGC~7582 was not in an obvious prolonged X-ray high-state 
after the optical event. In fact comparison between 
RXTE and {\it BeppoSAX} 
revealed a doubling of X-ray flux between 1998 October and November,
while the optical flux continued to decline by 30\% between 
1998 October 6 and 21 (A99).

In the context of the {\it BeppoSAX} model for the X-ray absorber, 
the broad optical permitted  lines must originate somewhere within
the full screen, which has sufficient opacity to hide those lines from view.  
The appearance of broad components on the permitted lines is 
 explained if holes appear in the full screen. This does not 
require the full screen to lie outside of the thick absorber, 
because 40\% of the line-of-sight is clear of the thick 
absorber, thus holes in the full screen would lead to some 
clear lines-of-sight to the broad-line-region.   
Holes or changes of opacity could occur if material forming the full screen is 
 inhomogeneous and is moving relative to the line-of-sight. 
If we attribute the optical changes in NGC~7582 to 
holes appearing in the full screen then 
we also expect measurable changes in optical reddening to be 
observed. This scenario has been previously suggested 
to explain similar changes in the optical spectra of 
Mrk~993 and Mrk~1018 (Goodrich 1989, 1995 respectively). 
A99 find the observed variations in the optical spectrum 
to be inconsistent with 
a simple decrease in optical reddening. However, as noted by A99, 
the gas and dust in 
NGC~7582 may result in a different reddening law than that assumed 
and 
even more importantly the starburst ``contamination'' of nuclear 
spectrum has to be removed, this has been done in the absence  
of high-spatial-resolution spectroscopy leaving 
some uncertainty in the optical reddening results. 
We feel these uncertainties leave open the possibility of 
a change in the full screen leading to the observed events.
 The soft-band X-ray flux appears highly variable on timescales of years, 
consistent with a model where the full screen  
allows a variable amount of continuum  leakage.
However, as noted above, the origin of the soft flux itself is 
ambiguous and there are other possible explanations for an increase of this
kind (variation in a spatially-unresolved 
source close to the nucleus, for example).

A99 discuss another explanation for the behavior of 
NGC~7582, suggesting the broad line components were 
produced in a  SN IIn explosion, these can produce  
broad lines without P-Cygni profiles (Schlegel, 1990). 
Starburst activity is important close to the nucleus of this 
source (Oliva et al 1995) and so a supernova 
event could occur so close to the nucleus as to be 
spatially unresolved from it, and could also produce the observed 
increase in soft X-ray flux. 
 A99 compare the evolution of the 
optical spectrum (widths of the broad line components) 
of NGC~7582 to that of the SN IIn SN 1988Z, 
finding good agreement. 
Our optical data  also lie on the A99 evolutionary curve (for line width).

\subsection{The Big Picture}
 
The detection of hard components, such as that found 
in the {\it BeppoSAX} data of NGC 7582,  are  
of great interest with regard to the question of the 
true intrinsic luminosity of type 2 Seyferts. Taking the {\it BeppoSAX} 
model, and calculating the {\it unabsorbed} 2--10 keV 
luminosity yields $L_{2-10}=10^{43} {\rm erg\ s^{-1}}$, 
a factor of several higher than the luminosity inferred from 
the {\it ASCA} measurements. As pointed out by 
Matt et al (1999) these 
hard components peak at a few tens of keV, the same energy as the
cosmic X-ray background. The spectra of many other 
Seyferts also appear flat when 
viewed in the 2-10 keV bandpass, e.g. NGC~2110 (Hayashi et al 1996; 
Malaguti et al. 1999), NGC~5252 (Cappi et al 1996), 
IRAS 04575-7537 (Vignali et al. 1998) and NGC~7172 (Guainazzi et al 1998). 
These flat spectra and the evidence for ``excess iron K edges'' in several 
NELGs  (Turner et al. 1997) indicates that complex 
absorption is a widespread phenomenon and the X-ray flux above 
10 keV is likely  significantly underpredicted for 
the Seyfert population when  
 extrapolating  the spectrum from lower energies. 
Thus {\it BeppoSAX} results from flat-spectrum Seyferts indicate that more
of the X-ray background could be attributed to Seyfert 2s 
and NELGs than previously thought.

\section{Conclusions}

A {\it BeppoSAX} observation of NGC~7582 reveals a previously unobserved 
hard X-ray spectral component visible above 10 keV. 
Correlated variability 
above and below 10 keV suggest that a single continuum 
component dominates the 2 -- 100 keV band. 
Comparison with {\it RXTE} data shows the 
flux in the 2 -- 10 keV band has varied by factors of several 
in the last few months, thus the  nuclear flux appears  
not to be correlated with the gradual decline noted in optical flux 
which followed the unusual optical event of July 1998. 
The 0.5 -- 2 keV flux has brightened by a factor $\sim 2$ 
since 1994, and this flux change seems most likely closely 
associated with the nucleus of NGC~7582. 
  
The $\sim 2$ -- 100 keV spectrum can be parameterized by a powerlaw of photon 
index $\Gamma=1.95$ which is 
significantly steeper ($\Delta \Gamma=0.40$) 
 than an {\it ASCA}  measurement of four years earlier. 
The X-ray absorber can be modeled 
as two distinct zones, a thick absorber with 
$N_H \sim 1.6 \times 10^{24}{\rm cm^{-2}}$ 
 covering $\sim 60\%$ of the nuclear source  
plus  a full-covering screen  
$N_H \sim 1.4 \times  10^{23}{\rm cm^{-2}}$. 
An absorber which is thick at the center and tapered 
towards the edges, and a distribution of clouds of differing 
column densities  are geometries which are strongly disfavored by the data.
The column density of the full screen appears to have increased 
by $\Delta N_H \sim 7 \times 10^{22}{\rm cm^{-2}}$ over the last 
four years. This result confirms the previous finding of column variability 
in this source (Xue et al 1998, Warwick et al. 1993). 

 It is possible that 
observed increases in soft X-ray flux over timescales of years 
are due to the appearance 
of holes in the full screen of the X-ray absorber, allowing 
a small leakage of the nuclear continuum (at the level of $\sim$ 1\% or less).
However, it is difficult to discern the relative contributions to the 
soft X-ray flux from the active nucleus and 
any outburst in the starburst regions of the host galaxy. Thus 
these data are also consistent with the suggestion by Aretxaga 
et al (1999) that the radiative onset of a type IIn supernova 
caused the observed change in properties of NGC~7582.

\section{Acknowledgements}
We thank the {\it BeppoSAX} satellite operations team. This research 
has made use of SAXDAS linearized and cleaned event files produced at 
the {\it BeppoSAX} Science Data Center. We also thank M. Cappi, 
F. Haardt for useful comments. Thanks to  Itziar 
Aretxaga and coauthors for providing their paper on the 
optical observations of NGC 7582 prior to publication, 
also to  Lorella Angelini for useful 
discussions about the data reduction, Steve Kraemer for 
bringing the IAU circular to our attention, Mike Crenshaw for
help with reduction of the optical data and the anonymous referee for
useful comments.

\newpage
\clearpage

\setcounter{figure}{0}
\begin{figure}
\plotfiddle{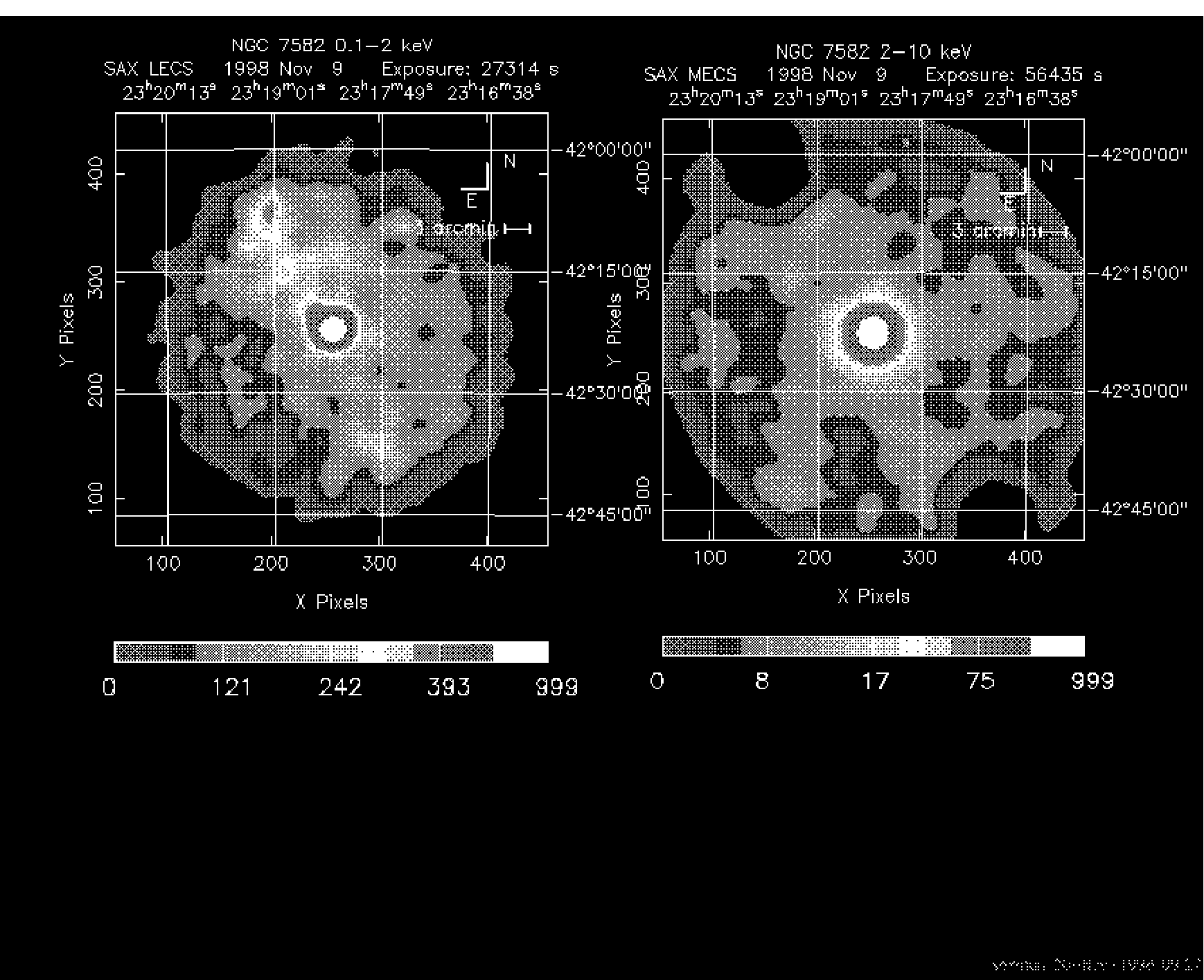}{10cm}{0}{50}{50}{-200}{0}
\caption{\it  The smoothed LECS and MECS images in the 0.1  -- 2 keV 
and 5-10 keV bands, respectively.  Extended emission and several 
nearby X-ray sources are evident in the LECS band, but none of 
the serendipitous sources provide significant flux above a few keV.
 }
\end{figure}
\clearpage

\begin{figure}
\plotfiddle{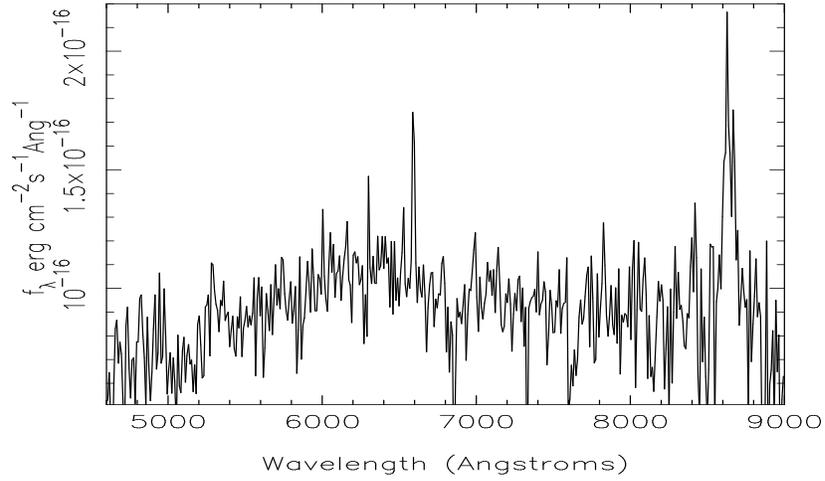}{5cm}{0}{65}{40}{-200}{0}
\caption{\it 
The optical spectrum of RX J231829.9-422041 which is $2'$ 
away from NGC~7582. Dropouts close to 
6860\AA\ and 8870\AA\ are instrumental effects due to charge traps in the 
CCD. The broad line observed at 8638\AA\ is likely to be 
H$\alpha$ at z=0.316, the line observed at 6593\AA\ is then 
identified as $[O {\sc iii}] \lambda 5007$. 
}
\end{figure}

\begin{figure}[h]
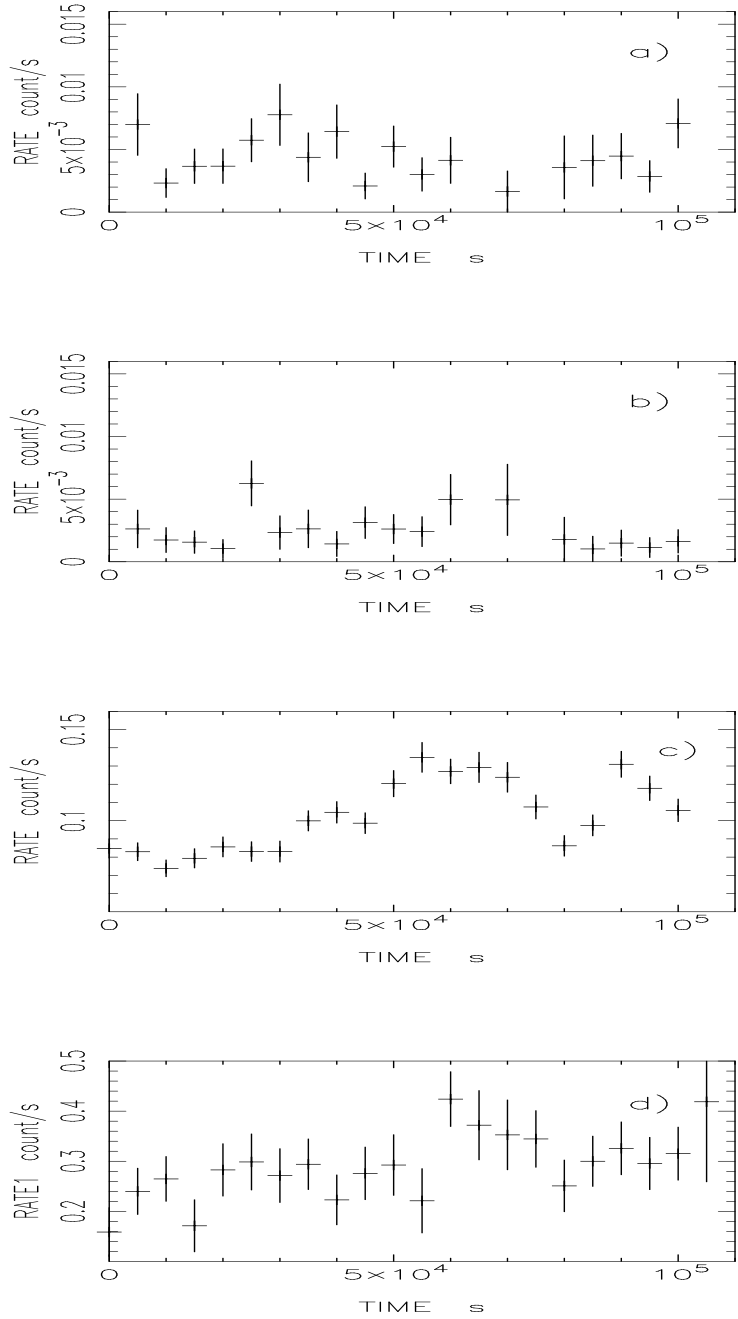

\plotfiddle{f3a.ps}{3cm}{0}{60}{25}{-200}{0}
\plotfiddle{f3b.ps}{3cm}{0}{60}{25}{-200}{-30}
\plotfiddle{f3c.ps}{3cm}{0}{60}{25}{-200}{-60}
\plotfiddle{f3d.ps}{3cm}{0}{60}{25}{-200}{-90}
\caption{\it The LECS source (a) and background (b) light curves in the 
0.1-1 keV band; c) MECS light curve from the 5-10 keV band and  
(d) the PDS light curve in the 13-60 keV band. All curves have 
5000s bins, the MECS background rate is negligible  and the PDS data 
have had the background level subtracted. }
\end{figure}

\begin{figure}[h]
\plotfiddle{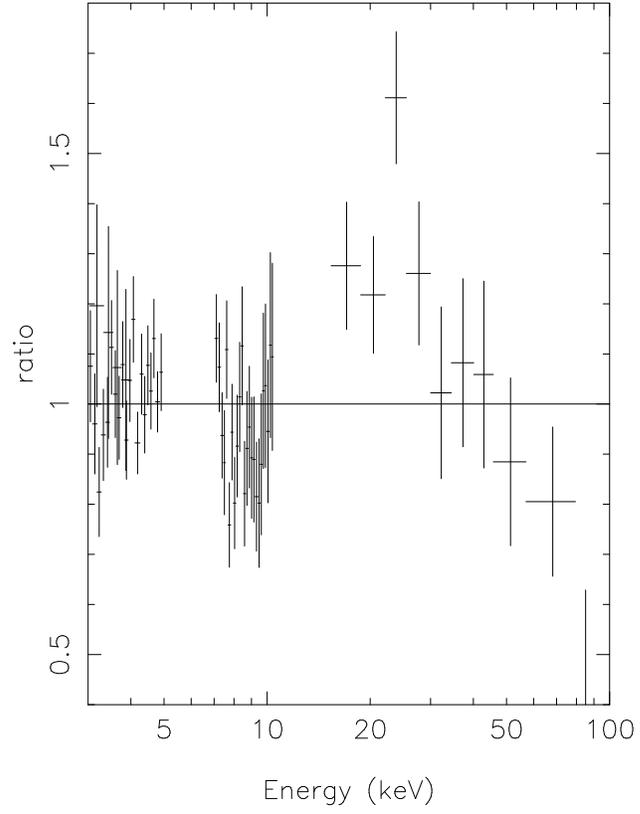}{8cm}{0}{50}{50}{-100}{0} 
\caption{\it The BeppoSAX  data from 3-5 keV plus 7-100 keV 
 compared to a simple model of an absorbed powerlaw.}
\end{figure}

\begin{figure}[h]
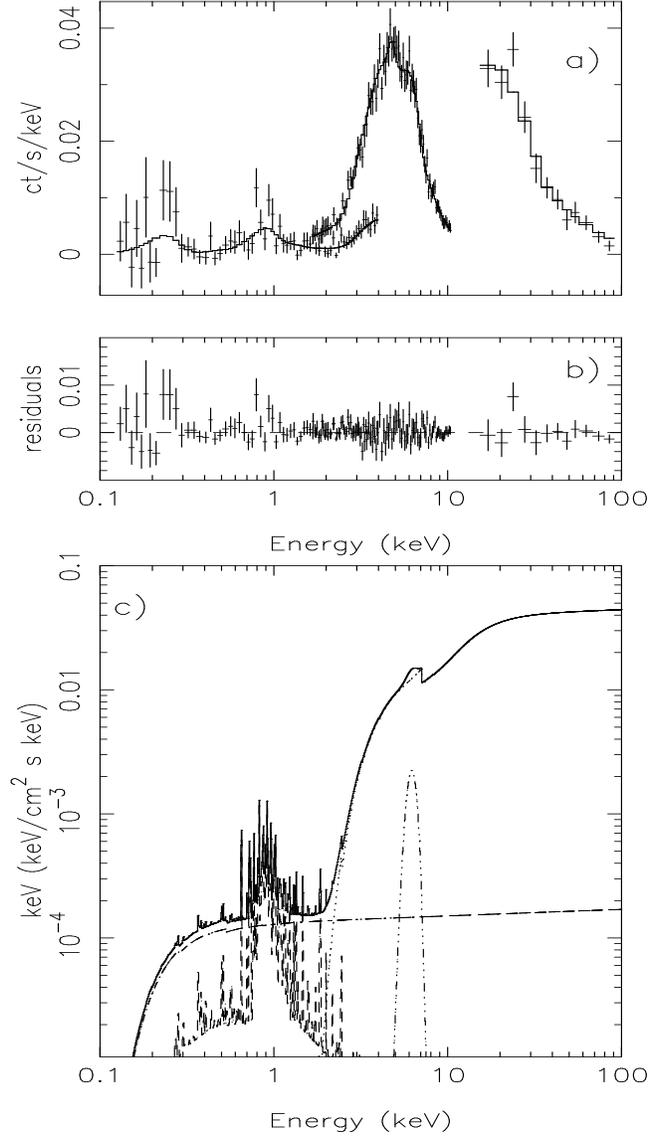

\plotfiddle{f5a_b.ps}{5cm}{0}{50}{35}{-120}{+25}
\plotfiddle{f5c.ps}{5cm}{0}{50}{35}{-120}{-45}
\caption{\it The BeppoSAX 
data (a) and residuals (b) compared to the best-fitting 
dual-absorber model (c). The solid line shows the sum of 
all components. The dotted line shows the powerlaw continuum, 
the turnover at $\sim$ 20 keV is due to the thick absorber and 
the turnover evident at a few keV is due to the full screen. 
The effects of the Galactic column become evident at $\sim 0.4 $ keV. 
The dashed line shows the {\sc mekal} component, 
the dash-dot lines show the iron K$\alpha$ emission 
line and  the unattenuated continuum. 
}
\end{figure}

\begin{figure}[h]
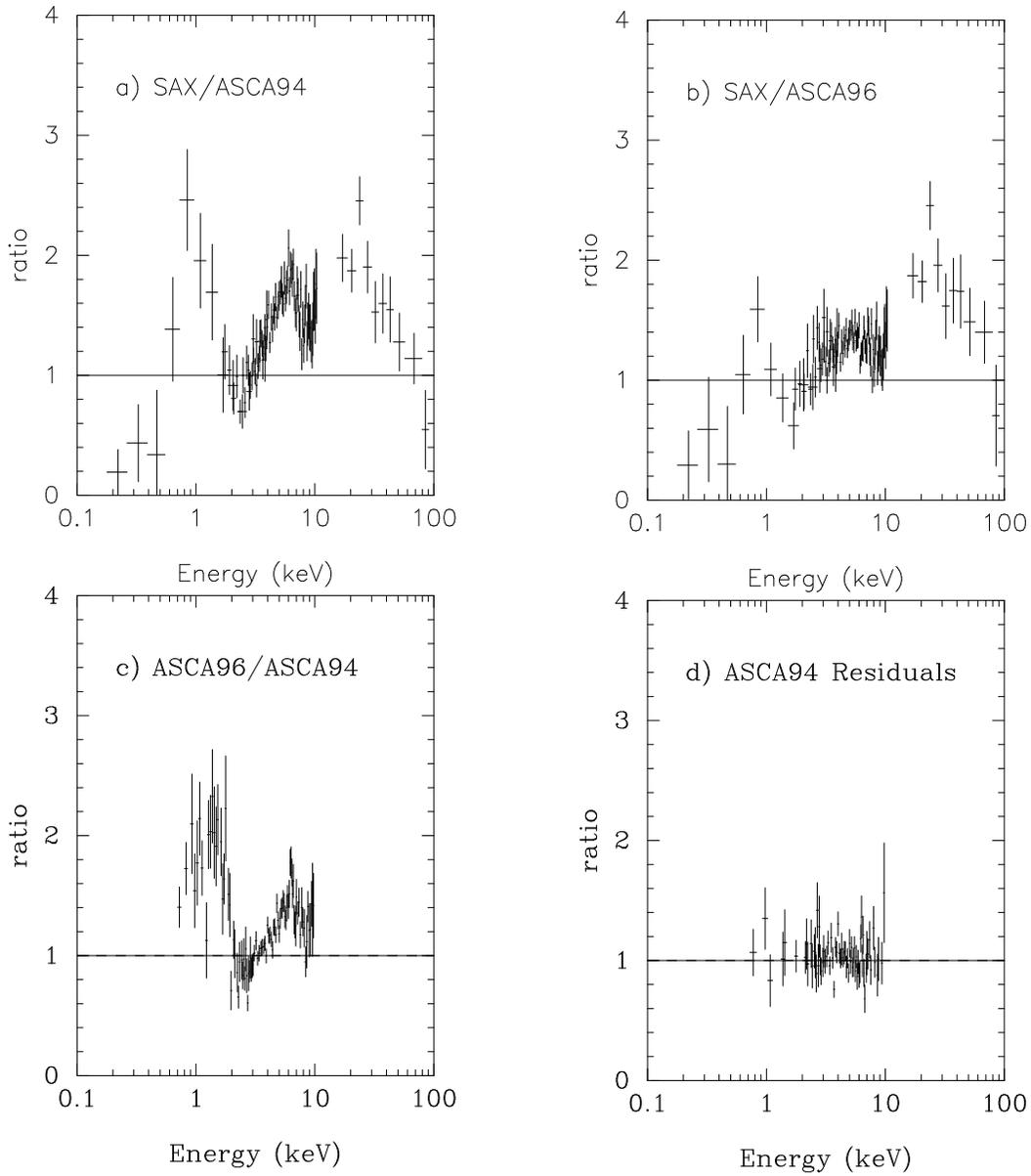

\plotfiddle{f6a.vps}{3cm}{0}{35}{35}{-240}{-100} 
\plotfiddle{f6b.vps}{3cm}{0}{35}{35}{-20}{0}
\plotfiddle{f6c.vps}{3cm}{0}{35}{35}{-240}{-120}
\plotfiddle{f6d.vps}{3cm}{0}{35}{35}{-20}{-20}
\caption{\it a) The BeppoSAX  data, compared to the model fitting  the 
 ASCA 1994 data; b) The  BeppoSAX data compared to the second  ASCA 
observation, from 1996; c) The ASCA data from 1996 compared to the 
model for the 1994 data; d) The ASCA  residuals after a fit to the 1994 data}
\end{figure}

\begin{figure}[h]
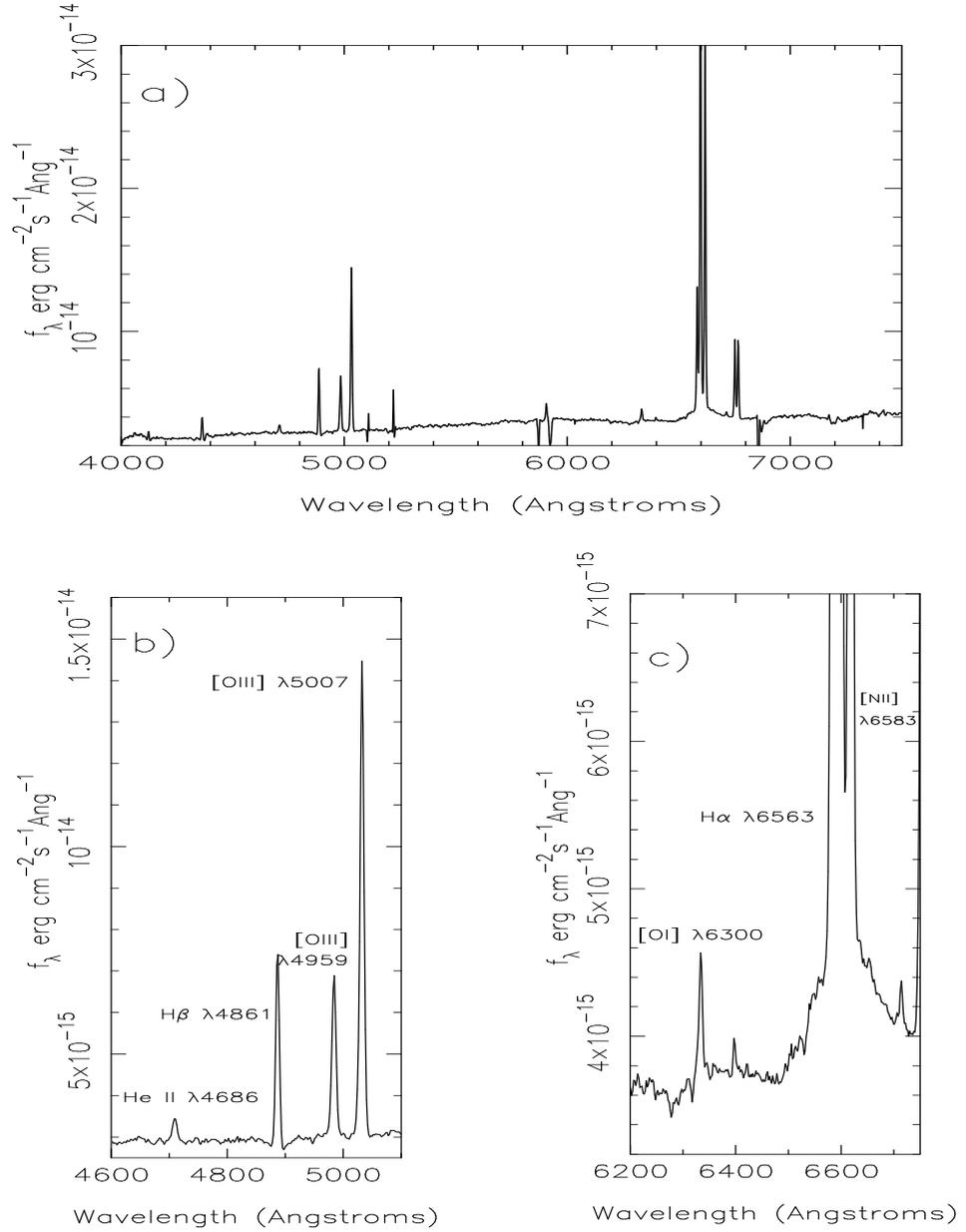

\plotfiddle{f7a.ps}{10cm}{0}{75}{40}{-210}{0}
\plotfiddle{f7b.ps}{5cm}{0}{65}{40}{-200}{-50}
\plotfiddle{f7c.ps}{5cm}{0}{65}{40}{-150}{+110}
\caption{\it a) The optical spectrum of NGC~7582 (z=0.0053) taken at
the CTIO 4m telescope; b) a close-up  of the H$\beta$ region 
and c) a closeup of the H$\alpha$ region }
\end{figure}

\end{document}